\documentclass[twocolumn,showpacs,amsmath,amssymb,prl]{revtex4}
\usepackage{graphicx}

\begin{document}
\title{Thermal diode: Rectification of heat flux}

\author{Baowen Li$^a$, Lei Wang$^b$, and Giulio Casati$^{c,d,a}$} 
\affiliation{$^a$ Department of Physics, National University of Singapore, Singapore 117542, Republic of Singapore\\
$^b$Temasek Laboratories, National University of Singapore, Singapore 119260, Republic of Singapore\\
$^c$International Center for Nonlinear and Complex Systems, 
Universita' degli studi dell'Insubria, Como\\
$^d$Istituto Nazionale di Fisica della Materia, Unita' di Como, Italy, and \\
Istituto Nazionale di Fisica Nucleare, sezione di Milano, 
Milano, Italy}
\date{Received 8 July 2004 and published online 27 October 2004}

\begin{abstract}
By coupling two nonlinear one dimensional lattices, we demonstrate a thermal diode model that works in a wide range of system parameters. We provide numerical and analytical evidence for the underlying mechanism which allows heat flux in one direction while the system acts like an insulator when the temperature gradient is reversed. The possible experimental realization in nano scale systems is briefly discussed.
\end{abstract}
\pacs{44.10.+i,  05.45.--a, 05.70.Ln, 66.70.+f}
\maketitle

The understanding of the underlying dynamical mechanisms which determines the macroscopic laws of heat conduction is a long standing task of non-equilibrium statistical mechanics\cite{Review}. Recent years have witnessed some important progress in this direction even though a satisfactory  understanding is, so far, unavailable. For example, it has been surmized that
for a one dimensional lattice,  momentum conservation leads, in general, to anomalous heat conduction\cite{Prosen}, with  heat conductivity $\kappa$ which diverges with the system size $L$ as $L^{2/5}$\cite{FPU}. If the transverse motion is considered, then $\kappa \sim L^{1/3}$\cite{WangLi04}. A connection between the anomalous conductivity and anomalous diffusion has been also established\cite{LiWang03}. Moreover, after two decades of debates, it is now clear that exponential local instability is not a necessary condition for the validity of Fourier law\cite{ChaosFL}.

A better understanding of the mechanism of heat conduction may also lead to potentially interesting applications based on the possibility to control the heat flow. Indeed, recently, a model for a thermal rectifier has been proposed\cite{diod}in which the rectifying effect is obtained by acting on the parameters which control the nonlinearity of the lattice. Although this model is far away from a realistic implementation, nevertheless it opens the possibility to propose thermal devices which may have practical relevance.

In this Letter we demonstrate, via computer simulations, the possibility to build a thermal diode by coupling 
two nonlinear lattices. Indeed our model allows the heat flow from one end to the other, but it inhibits the flow in the opposite direction. The model works in a wide range of system parameters and the underlying microscopic mechanism is different from the one in ref\cite{diod} even though it is based on the same general idea. In the present case the ratio of the heat currents in the two opposite directions is 100 times larger than that in model\cite{diod} and it is therefore sufficiently large to encourage experimental work.

Our system consists of two segments of nonlinear lattices coupled together by a harmonic spring with constant strength $k_{int}$. 
Each segment is described by the Hamiltonian: 
\begin{equation}
H =\sum \frac{p^2_i}{2m}+\frac{1}{2}k(x_i-x_{i+1}-a)^2-\frac{V}{(2\pi)^2}\cos 2\pi x_i.
\label{eq:Ham}
\end{equation}
 
In fact, Eq. (\ref{eq:Ham}) is the Hamiltonian of the Frenkel-Kontorova (FK) model which is known to have normal heat conduction \cite{HLZ98}. 

For simplicity we set the mass of the particles and the lattice constant $m=a=1$. Thus the adjustable parameters are $(k_L,k_{int},k_R,V_L,V_R,T_L,T_R)$, 
where the letter L/R indicates the left/right segment and $T_{L,R}$ is the temperature of the left/right heat bath. In order to reduce the number of adjustable parameters, we set $V_R=\lambda V_L$, $k_R=\lambda k_L$, $T_L=T_0(1+\Delta), T_R=T_0(1-\Delta)$ and, unless otherwise stated, we fix  $V_L=5$, $k_L=1$  so that the adjustable parameters are reduced to four, $(\Delta, \lambda, k_{int}, T_0)$. Notice that when $\Delta>0$, the left bath is at higher temperature and vice versa when $\Delta<0$. 

In our numerical simulations we use fixed boundary conditions and  
the $N$ particle chain is coupled, at the two ends, with heat baths at temperatures $T_L$  and $T_R$ respectively.
We use Langevin heat baths and we have checked that our results do not depend on the particular heat bath realization (e.g. Nos\'e-Hoover heat baths).
We then integrate the differential equations of motion by using the 5th-order Runge-Kutta algorithm as described in \cite{RKBcode}.
We compute  the temperature profile inside the system, i.e.\
the local temperature at site $n$ defined as  
$T_n = m \langle \dot{x_n}^2 \rangle $,   
where $\langle \hspace{0.2cm}
\rangle$ stands for temporal average, and the local heat flux 
$J_n = k  \langle \dot{x_n} (x_n - x_{n-1})\rangle$ \cite{Review}.
The simulations are performed long enough to allow the system to reach a
non-equilibrium steady state where the local  heat flux
is constant along the chain.

In Fig. \ref{fig:jdelta} we plot the heat current $J$ versus $\Delta$  for different temperatures $T_0$. It is clearly seen that when $\Delta>0$ the heat current increases with $\Delta$, while in the region  $\Delta<0$ the heat current is almost zero, i.e. the system behaves as a  thermal insulator. The results in Fig. \ref{fig:jdelta} show that our model has the rectifying effect in a wide range of temperatures. The rectifying efficiency depends on temperature as well as on other parameters as described below:
 
\begin{figure}
\includegraphics[width=\columnwidth]{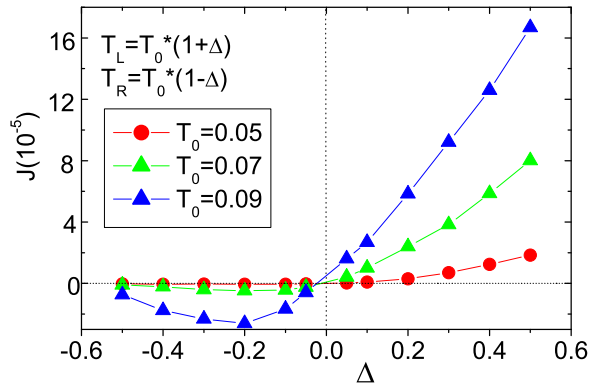}
\vspace{-1.2cm}
\caption{\label{fig:jdelta}Heat current $J$ versus the dimensionless temperature difference $\Delta$ 
for different values of $T_0$. Here the total number of particles $N= 100$, $k_{int}=0.05$, $\lambda=0.2$. The lines are drawn to guide the eye.}
\end{figure}

{\it $k_{int}$ effect}. The interface elastic constant $k_{int}$ is a very important parameter as it plays the role of coupling left and right lattices. By adjusting this parameter one can control the heat flow. Indeed, once the parameters of the two lattices are fixed, then the smaller the coupling the smaller is the heat current through the system. 

In order to describe quantitatively the rectifier efficiency we introduce the ratio, $|J_+/J_-|$ where $J_+$ is the heat current (from left to right) when the bath at higher temperature $T_+$ is at the left end of the chain and $J_-$ is the heat current (from right to left) when the left end of the chain is in contact with the bath at lower temperature $T_-$.

Fig.\ref{fig:jkint}(a) shows that, by varying $k_{int}$, the rectifier efficiency $|J_+/J_-|$ changes from about two times at $k_{int}=0.5$ to more than 100 times for $k_{int} \le 0.01$. More importantly, this figure shows that the rectifying effect is very significant in a wide range of $k_{int}$. In the inset of this figure we show $|J_{\pm}|$ versus $k_{int}$ for  a system of eight particles. It is seen that, for small $k_{int}$, numerical data follow the dependence $J_{\pm} \propto k^2_{int}$  over almost two orders of magnitude. Therefore the ratio $|J_+|$ / $|J_-|$ is a constant independent of $k_{int}$. 

\begin{figure}
\includegraphics[width=\columnwidth]{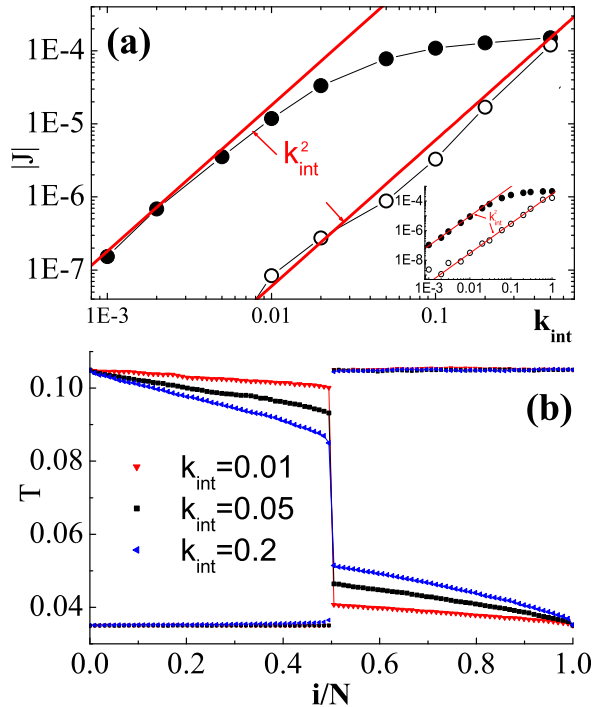}
\vspace{-1.2cm}
\caption{\label{fig:jkint}(a) Heat current $|J_{\pm}|$ versus the interface elastic constant $k_{int}$ for $N=100$, $\lambda=0.2$, $T_+=0.105$, $T_-=0.035$. The solid circles are the current $J_+$ while open circles are the current $J_-$. Inset in (a) is $|J_{\pm}|$ versus $k_{int}$ for a system of eight particles. The solid lines in the figure and in the inset have slope equal two. (b) The temperature profile for $k_{int}=0.01, 0.05$, $0.2$ for the same parameters of \label{fig:jkint}(a)}
\end{figure}

Fig. \ref{fig:jkint}(b) shows the temperature profile for different $k_{int}$. There exists a large temperature jump at the interface and this jump is much larger when the high temperature bath is at the right end. In this case there is very small temperature gradient inside each lattice and, as a consequence, the heat current is almost vanishing.

\begin{figure}
\includegraphics[width=\columnwidth]{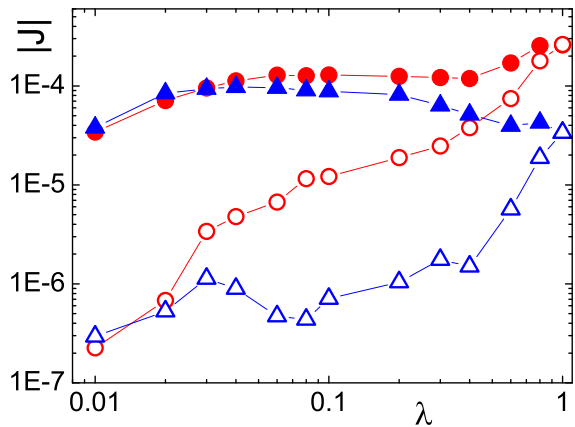}
\vspace{-1.2cm}
\caption{\label{fig:jlambda}Heat current versus $\lambda$ for $k_{int}=0.2$ (circles), and $k_{int}=0.05$ (triangles). The solid symbols refer to $T_L=T_+, T_R=T_-$,  while empty symbols refer to  $T_L=T_-, T_R=T_+$. Here $T_+=0.105$, $T_-=0.035$ and $N=100$.  }
\end{figure}

{\it Effect of the lattice parameter $\lambda$.}
As it is known \cite{HLZ98}, in the FK model the elastic constant and the strength of the on-site potential can be scaled to a single parameter. Therefore it is sufficient to study the properties of the system (1) as a function of the single parameter $\lambda$.

In Fig. \ref{fig:jlambda}, we show the current $J_{\pm}$ versus $\lambda$ for two different interface constants $k_{int}=0.05$, and $k_{int}=0.2$. This figure clearly show that in a wide range of parameters our model has a quite good  rectifying efficiency, i.e. $|J_+/J_-| \sim 100$.

\begin{figure}
\includegraphics[width=\columnwidth]{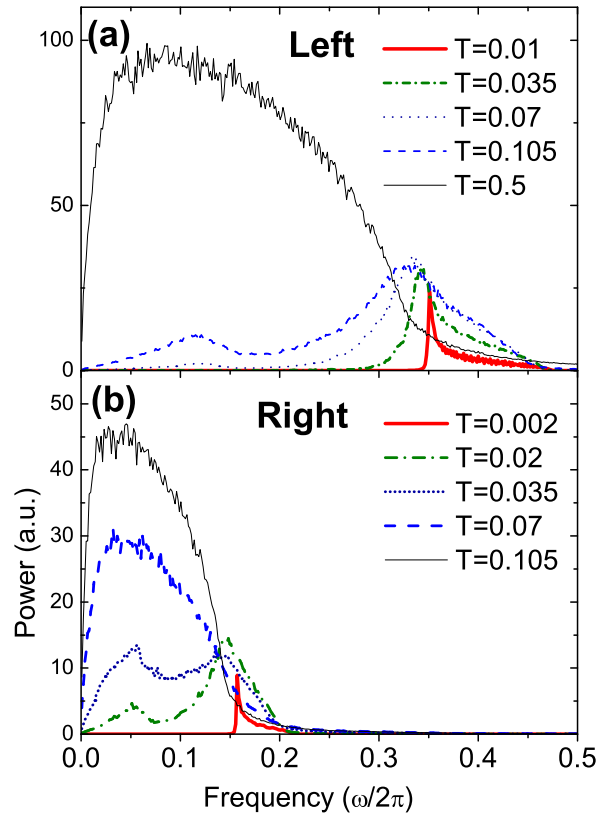}
\vspace{-1cm}
\caption{\label{fig:spectrum}Spectra of the two particles at the interface for different temperatures at $k_{int}=0$. (a) particle at the left side of the interface, (b) particle at the right side of the interface. Here $\lambda=0.2$, $N=100$.}
\end{figure}

{\it Rectifying mechanism.} 
To understand the underlying rectifying mechanism, let's start from the energy spectrum of the interface particles. Fig. \ref{fig:spectrum} shows the phonon spectra of the left and right interface particles at different temperature when the two lattices are decoupled ($k_{int}=0$). 

The match/mismatch of the energy spectra of the two interface particles controls the heat current. It is clearly seen from Fig.\ref{fig:spectrum} that, if the left end is in contact with the high temperature bath $T_L$, and the right end with the low temperature bath $T_R$ ($<T_L$), then the phonon spectra of the two particles at interface overlap in a large range of frequencies, thus the heat current can easily go through the system from the left end to the right end. However, if  the left end is at lower temperature   $T_L$ and the right end is  at higher temperature $T_R$ ($>T_L$), then the phonon spectrum of the right interface particle is mainly in the low frequency part, while the left interface particle is in the high frequency part. Then there is almost no overlap in phonon frequency, and the heat current can hardly go through from right to left, and the system behaves as an insulator. Why the left and right particles at the interface have so different phonon spectra? This can be understood from the following analysis in different temperature regimes.

(1) {\it Low temperature limit}. 
At low temperature, the particle is confined in the valley of the on-site potential. By linearizing the equation of motion one can easily obtain the frequency band\cite{diod}: 
\begin{equation}
\sqrt{V}<\omega <\sqrt{V+4k}.
\label{eq:lowfre}
\end{equation}
For  the case of Fig.\ref{fig:spectrum} with T=0.01 (left) and T=0.002 (right), this corresponds to $0.36<\omega/2\pi<0.48$ for the left particle and to $0.16<\omega/2\pi<0.21$ for the right particle. 

As the temperature is increased, the interparticle potential $kx^2/2$ becomes more and more important until a critical value  $T_{cr}\approx V/(2\pi)^2$ is reached (we take the Boltzman constant equal unity), when the kinetic energy is large enough to overcome the on-site potential barrier. At this point low frequency appears and this happens at the critical temperatures
$T_{cr}=0.13$ for $V=5$ (left), and $T_{cr}=0.025$ for $V=1$ (right). This is in quite good agreement with the data of Fig. \ref{fig:spectrum}. 

(2) {\it High temperature limit}.
In the high temperature limit the on-site potential can be neglected, the system is close to two coupled harmonic chains, and the phonon band is \cite{kittle},
\begin{equation}
0<\omega<2\sqrt{k},
\label{eq:highfre}
\end{equation}
which gives $0<\omega/2\pi<0.32$ for the left particle and  $0<\omega/2\pi<0.14$ for the right particle, again in good agreement with Fig. \ref{fig:spectrum}.

In fact, in order to optimize the rectifying effect, one should avoid the overlapping of the phonon bands in the low temperature limit (Eq.\ref{eq:lowfre}) and that in the high temperature limit (Eq.\ref{eq:highfre}) for each segment of the system. According to the above estimates, one should have $V>4k$, which is satisfied for the case of Fig. \ref{fig:spectrum}. 

The analysis presented here allows also to understand why in Fig.\ref{fig:jdelta} there is a region of negative conductivity for $T_0 = 0.09$ and $-0.5<\Delta<-0.2$. Indeed in this interval it is seen that by increasing the temperature difference between the two heat baths, the current decreases! As explained above, the reason for such unusual behavior is due to the fact that by decreasing the temperature of the left bath, the phonon spectra of the two interface particles separate one from the other.

It is worth pointing out that it is the temperature dependence of the phonon band that makes the rectifying effect possible. This dependence is due to the nonlinearity of the potential and therefore it should be possible to observe the rectifying effect in any nonlinear lattice. The difference between different choices of the on-site potential is quantitative rather than qualitative. For example, we have observed the rectifying effect in the $\phi^4$ model wherein the on-site potential is $kx^2/2+\beta x^4/4$, but the rectifying efficiency is not as significant as it is in the FK model. The main reason for the very large rectifying efficiency in the FK model is the finite height of the on-site potential.

\begin{figure}
\includegraphics[width=\columnwidth]{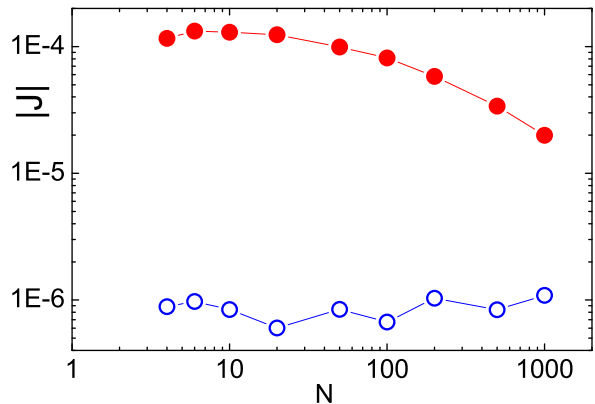}
\vspace{-1.2cm}
\caption{\label{fig:jvsn} The finite size effect of the rectifying efficiency. Here we plot $|J_+|$ (solid circles) and $|J_-|$ (open circles) versus $N$. Parameters are $\lambda=0.2, k_{int}=0.05$, $T_+=0.105$, and $T_-=0.035$.}
\end{figure}

{\it Finite size effect}.  The results shown above are for a system with total number of particles $N=100$.
Since the rectifying mechanism in our model is due to the coupling between two dissimilar lattices it is reasonable to expect that the number of nonlinear oscillators will definitely affect the rectifying efficiency. This is shown in Figure \ref{fig:jvsn} where we plot $|J_+|$ and $|J_-|$ versus the system size $N$ with $T_+=0.105$, and $T_-=0.035$.

The decrease of the rectifying efficiency by increasing $N$ (Fig\ref{fig:jvsn}) and $k_{int}$ (Fig\ref{fig:jkint}) can be understood in the following way. If we denote by $R$ the sum of thermal resistances of the two segments, then the total resistance of the system is 
$R_{\pm}=R+R^{\pm}_{int}$, where $R^{\pm}_{int}$ is interface resistance when $\Delta>0$ and $\Delta<0$, respectively.
As  $J_{\pm} = (T_L -T_R)/R_{\pm}$, the rectifying efficiency is $|J_+/J_-| \approx (R+R_{int}^-)/(R+R^+_{int})$.
Notice that $R$ increases with $N$ while $R^{\pm}_{int}$ are $N$ independent, they decrease with increasing $k_{int}$, and  $R^+_{int}$ is much smaller than $R^-_{int}$. It is therefore clear that the efficiency decreases with increasing 
$k_{int}$ (Fig\ref{fig:jkint}) at fixed $N$ and by increasing $N$ at fixed $k_{int}$ (Fig\ref{fig:jvsn}).

For the data of  Fig.\ref{fig:jvsn} we have $R_{int}^- \gg R$, and therefore 
$J_-$ is a small constant that is almost independent of $N$, while $J_+\sim 1/N$. As $N$ is increased over a critical value $N_c$ where $|J_+| \approx |J_-|$, the rectifying effect vanishes. For the case of Fig. \ref{fig:jvsn} this happens at $N_c \sim 10^5$. However, we should point out that the value of $N_c$ depends on $k_{int}$ and by decreasing $k_{int}$ we may increase $N_c$.

Finally, we would like to discuss the possibility of an experimental realization of our diode model. In our calculations, we take dimensionless units. For a typical atom, the dimensionless temperature $T_0$ used in this paper is related to the real temperature $T^r$ by $T^r \sim 10^2-10^3 T_0$ \cite{HLZ98}. Therefore $T_0 \sim 10^0$ is of the order of room temperature. If we consider the lattice distance of 1 $\AA$, then a lattice of 100 particles is about 10 nanometers long, which is of a size scale that can be fabricated in nowaday's laboratory. Another problem for  real experiments is that the on-site potential is due to the interaction with a substrate. In our numerical simulations, we neglect the heat current in the substrate. This may affect the rectifying efficiency of the diode. Therefore, one needs to choose a substrate with substantially low thermal conductivity. 

In summary, we have devised a thermal diode by using two coupled nonlinear lattices. Our model exhibits a very significant rectifying effect in a very wide range of system parameters. The underlying mechanism has been discussed as well as a possible realization of the model in a nano scale system.

BL is
supported in part by Faculty Research Grant of National University of Singapore and the Temasek Young Investigator Award of DSTA Singapore under Project Agreement POD0410553. LW is supported by DSTA
Singapore under Project Agreement POD0001821. GC is partially supported by EU Contract No. HPRN-CT-2000-0156 (QTRANS) and by MURST (Prin 2003, Ordine e caos nei sistemi estesi non lineari: strutture, stocasticita' debole e trasporto anomalo).

\end{document}